\documentclass[a4paper]{article}

\usepackage{ISCSLP2022}
\usepackage{multirow}
\usepackage{enumitem}
\title{The NPU-ASLP System for The ISCSLP 2022 Magichub Code-Swiching ASR Challenge}
\name{Yuhao Liang, Peikun Chen, Fan Yu, Xinfa Zhu, Tianyi Xu, Lei Xie}
\address{
  Audio, Speech and Language Processing Group (ASLP@NPU), School of Computer Science,\\ Northwestern Polytechnical University, Xi’an, China 
  }
\email{liangyuhao@mail.nwpu.edu.cn, lxie@nwpu.edu.cn}

\begin{document}

\maketitle

\begin{abstract}
\vspace{-0.2cm}
    This paper describes our NPU-ASLP system submitted to the ISCSLP 2022 Magichub Code-Switching ASR Challenge. In this challenge, we first explore several popular end-to-end ASR architectures and training strategies, including bi-encoder, language-aware encoder (LAE) and mixture of experts (MoE). To improve our system’s language modeling ability, we further attempt the internal language model as well as the long context language model. Given the limited training data in the challenge, we further investigate the effects of data augmentation, including speed perturbation, pitch shifting, speech codec, SpecAugment and synthetic data from text-to-speech (TTS). Finally, we explore ROVER-based score fusion to make full use of complementary hypotheses from different models. Our submitted system achieves 16.87\% on mix error rate (MER) on the test set and comes to the 2nd place in the challenge ranking.
\end{abstract}


\vspace{-0.3cm}
\section{Introduction}
\vspace{-0.1cm}

Code-switching occurs when a speaker alternates between two or more languages. With fast globalization and frequent culture exchange, code-switching has become a common language phenomenon which poses significant challenges to speech and language processing tasks including automatic speech recognition (ASR). Code-switching may occur in the middle of a sentence (\textit{intra-sentential}) or at the sentence boundaries (\textit{inter-sentential}) while the former is considered to be more difficult to a speech recognizer. To promote reproducible research of Mandarin-English code-switching ASR, ISCSLP2022 has specifically held the Magichub Code-Switching ASR challenge$\footnote{https://magichub.com/competition/code-switching-asr-challenge}$, which provides a sizeable corpus and a common test-bed to benchmark the code-switching ASR performance.

Code-switching ASR has been explored for quite a long time since the conventional hybrid ASR paradigm~\cite{guo2018study}. Progress has also been advanced with several challenges specifically focusing on the code-switching phenomena~\cite{shi2020asru,shah2020first,diwan2021multilingual}.
With the recent advances in deep learning, neural end-to-end (E2E) frameworks, such as attention encoder decoder (AED)~\cite{vaswani2017attention,chorowski2015attention} and neural transducer~\cite{graves2012sequence}, have emerged as the mainstream for ASR with simplified system building pipeline and substantial performance improvement. However, modeling multiple languages simultaneously in a unified neural architecture is non-trivial because different languages (e.g., Mandarin and English) have significant differences in many aspects including modeling units and manner of articulation.

Recently, language-expert modules~\cite{lu2020bi, tian2022lae,hou2021exploiting} were proposed for modeling different languages by separated parameters in multilingual or cross-lingual settings, which can capture language-specific knowledge space effectively and mitigate overfitting caused by the poverty of code-switching data.

Specifically, network parameters were decomposed into language-specific parts (or \textit{experts}) in a bi-encoder structure, where each transformer encoder represents a language (i.e., Mandarin and English)~\cite{lu2020bi, tian2022lae}.

Meanwhile, the bi-encoder architecture can effectively leverage rich monolingual data from both languages.
But due to the lack of interaction between the separated encoders, the language-common feature space is apparently ignored. 
Therefore, language-aware encoder (LAE)~\cite{tian2022lae} was further proposed to address this problem by sharing the preliminary blocks before the language-specific experts, which could model both language-specific and language-common feature efficiently.
Instead of sharing only the preliminary blocks, mixture of experts (MoE)~\cite{hou2021exploiting} was designed to share the majority of parameters, which may be able to learn more language-common feature and be better suited to limited training data conditions.


Another difficulty is the data sparsity problem. As the language switch can be occurred anywhere in an utterance for the more difficult intra-sentential switch, it is hard to collect enough code-switching data and prediction of the switching position is rather difficult.

To overcome this problem, data augmentation might be a feasible solution, including text-to-speech (TTS) augmentation and text data augmentation.
Note that using synthetic data directly often has negligible gain or even misguides the ASR system because of the mismatch between the synthetic and real data. 

For better use of synthetic data, some additional loss functions~\cite{chen2022tts4pretrain,chen2021injecting} and filtering strategies~\cite{park2022unsupervised,hu2022synt++} were proposed to enforce the consistency of hypothesized labels between real and synthetic data.
For text augmentation, a machine translation model was usually adopted to expand on the original code-switching text~\cite{liu2021code}.

In this challenge, we approach the Mandarin-English code-switching ASR by exploration of both multi-lingual neural architectures and data augmentation. Specifically, we study the bi-encoder, LAE and MoE architectures reviewed above under the popular Conformer based AED framework implemented with two popular ASR toolkits -- ESPNet~\cite{watanabe2018espnet} and WeNet~\cite{yao2021wenet}. Various data augmentation methods, including speed perturbation, pitch shifting, audio codec augmentation, spectrum augmentation as well as text-to-speech augmentation. Specifically for TTS augmentation, a consistency loss~\cite{chen2022tts4pretrain} is proved to be effective for mitigating the mismatch in the distribution of real and synthetic data. We further explore the effectiveness of language modeling, including both internal language model as well as long context language model~\cite{wei2021context}. Finally, ROVER~\cite{fiscus1997post} is adopted for fusion of multiple hypothesis from various models, which has previously proven to be effective~\cite{yu2022summary,sun2018multiple,wang2022sjtu}. Our fusion system has achieved the lowest MER of 16.87\% on the test set, leading our submission to the 2nd place in the challenge.

\vspace{-0.2cm}
\section{Data Preparation\label{sec2}}
\vspace{-0.1cm}
In this section, we will introduce the datasets used in this challenge and augmentation methods applied to the training data.

\vspace{-0.3cm}
\subsection{Datasets\label{sec21}}
\vspace{-0.1cm}
The training data consists of two parts, one is the TALCS corpus~\cite{li2022talcs}, which is recorded from online one-to-one teacher-student English teaching. It contains roughly 587 hours of speech sampled at 16 kHz. The other part is MagicData-RAMC~\cite{yang2022open}, which contains 180 hours of conversational speech data recorded from native speakers of Mandarin Chinese over mobile phones with a sampling rate of 16 kHz.
\vspace{-0.3cm}
\subsection{Data Augmentation\label{sec22}}
\vspace{-0.1cm}
Data augmentation plays an important role in model training especially in low resource or limited data scenario and it can significantly increase data diversity and quantity. Our data augmentation methods are as follows.
\vspace{-0.1cm}
\begin{itemize}[leftmargin=*]
    \item \textbf{Speed perturbation}: Speed perturbation is the process of changing the speed of an audio signal without affecting its pitch. We apply it on all training data with speed factor 0.9, 1.0 and 1.1.
    \vspace{-0.1cm}
    \item \textbf{Pitch shifting}: Pitch shifting can effectively vary the timbre of the sound to increase data diversity. We use the \emph{Sox} audio manipulation tool to perturb the pitch in the range [-40, 40]. 
    \vspace{-0.1cm}
    \item \textbf{Audio codec}: We use an audio codec strategy to increase the diversity of voice without introducing external data. We use the FFmpeg tool in Opus format at 256kbps.
    \vspace{-0.1cm}
    \item \textbf{Spectrum augmentation}: To mitigate possible model over-fitting, the SpecAugment~\cite{park2019specaugment} method is applied to input features on every mini-batch. The SpecAugment includes time warping, masking on frequency channels, and masking of time steps. We utilized all of them during our training.
    \vspace{-0.1cm}
    \item \textbf{Text-to-speech (TTS)}: We adopt the structure of DelightfulTTS~\cite{Liu2021DelightfulTTS}, removing the prosody encoder and predictor and replacing the HiFiNet with HiFi-GAN V1~\cite{Kong2020HiFiGANGA}. We use style-id to distinguish the TALCS dataset and MagicDataRAMC dataset and build a multi-speaker multi-style TTS system. The TTS system takes text and style-id as input. Moreover, the speaker representation is extracted by GST~\cite{Wang2018StyleTU} and added to the variance adaptor output. As the code-switching data do not have speaker id, we make a semi-supervised classification constraint on the speaker representation. Specifically, the speaker classifier loss of code-switching data is not calculated, and the model will softly determine which speaker each utterance contains. Thus, we can synthesize code-switching audio with dialogue style with the mel-spectrum of the target speaker. Specifically, We select 20 hours of data from the training data to train the TTS model. After training, we synthesize 100 hours of data using the transcript of  TALCS and the style-id of MagicDataRAMC.
\end{itemize}
\vspace{-0.4cm}

\begin{figure}[h]
	\includegraphics[width =0.95\linewidth]{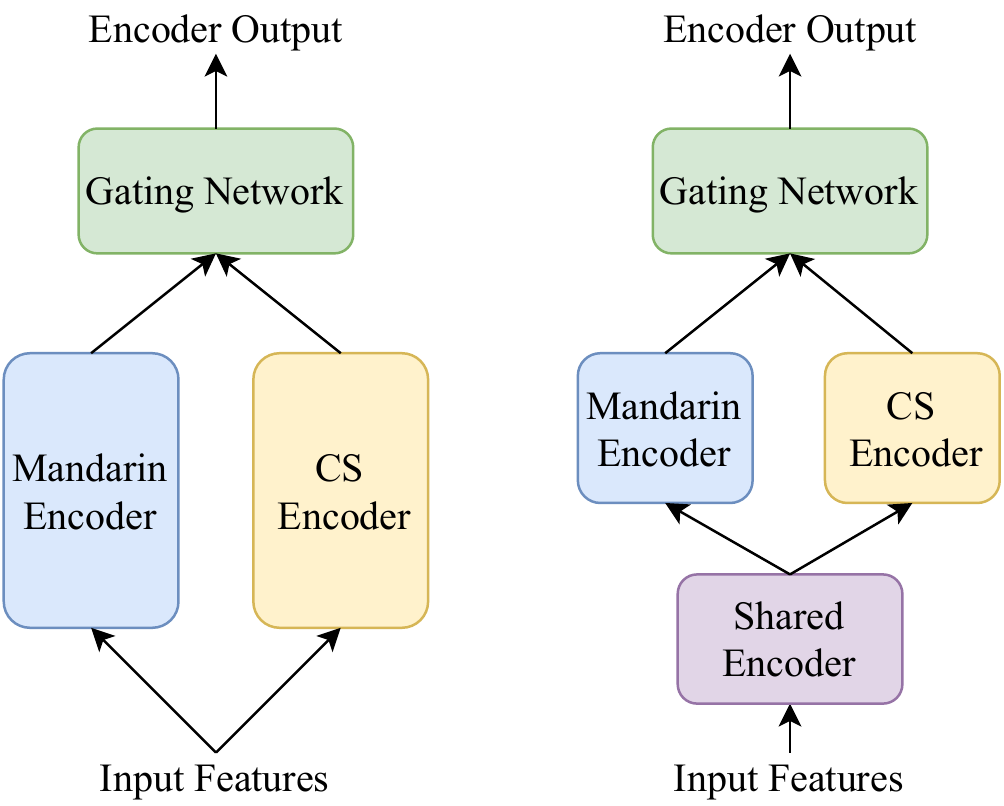}
	\caption{
	The architecture of bi-encoder (left) and LAE (right), CS indicates code-switching.
	}
	\label{biencoder}
\vspace{-16pt}
\end{figure}
\section{System Description\label{sec3}}
\vspace{-0.1cm}
In this section, we present our ASR approaches explored in this challenge. Our systems mainly consist of acoustic models, language models and model fusion. As for acoustic models and language models, various architectures and training strategies are utilized.
\vspace{-0.3cm}
\subsection{Acoustic Model\label{sec31}}
\vspace{-0.1cm}
For acoustic models, we try three variants of the encoder which can lead to various outputs. This variety will offer benefits to model fusion.
\vspace{-0.3cm}
\subsubsection{Bi-encoder\label{sec311}}
\vspace{-0.1cm}
As shown on the left of Figure \ref{biencoder}, bi-encoder~\cite{lu2020bi} network uses two separate encoders to serve as the Mandarin expert and the English expert for better capturing language-specific knowledge. Bi-encoder produces two different views of input code-switching audio and integrates by a gating network on frame level. Due to limited data in this challenge, no monolingual English data is available and outside data is not allowed to use. Hence, the Mandarin encoder is pre-trained on MagicData-RAMC and the English encoder is pre-trained on TALCS which is bilingual. With initialized encoders, the model is finetuned on all data.
\vspace{-0.3cm}
\subsubsection{Language-Aware Encoder\label{sec312}}
\vspace{-0.1cm}
Similar to bi-encoder, language-aware encoder (LAE)~\cite{tian2022lae} also contains two parallel expert blocks. As shown on the right of Figure \ref{biencoder}, The main difference between LAE and bi-encoder is that LAE contains an extra shared block followed by the expert blocks. LAE can be explained as a variant of bi-encoder that shares the parameters of primary layers. Shared blocks provide more genetic features in preliminary encoding.
\vspace{-0.3cm}
\subsubsection{Mixture of Expert\label{sec313}}
\begin{figure}
	\includegraphics[width =0.80\linewidth]{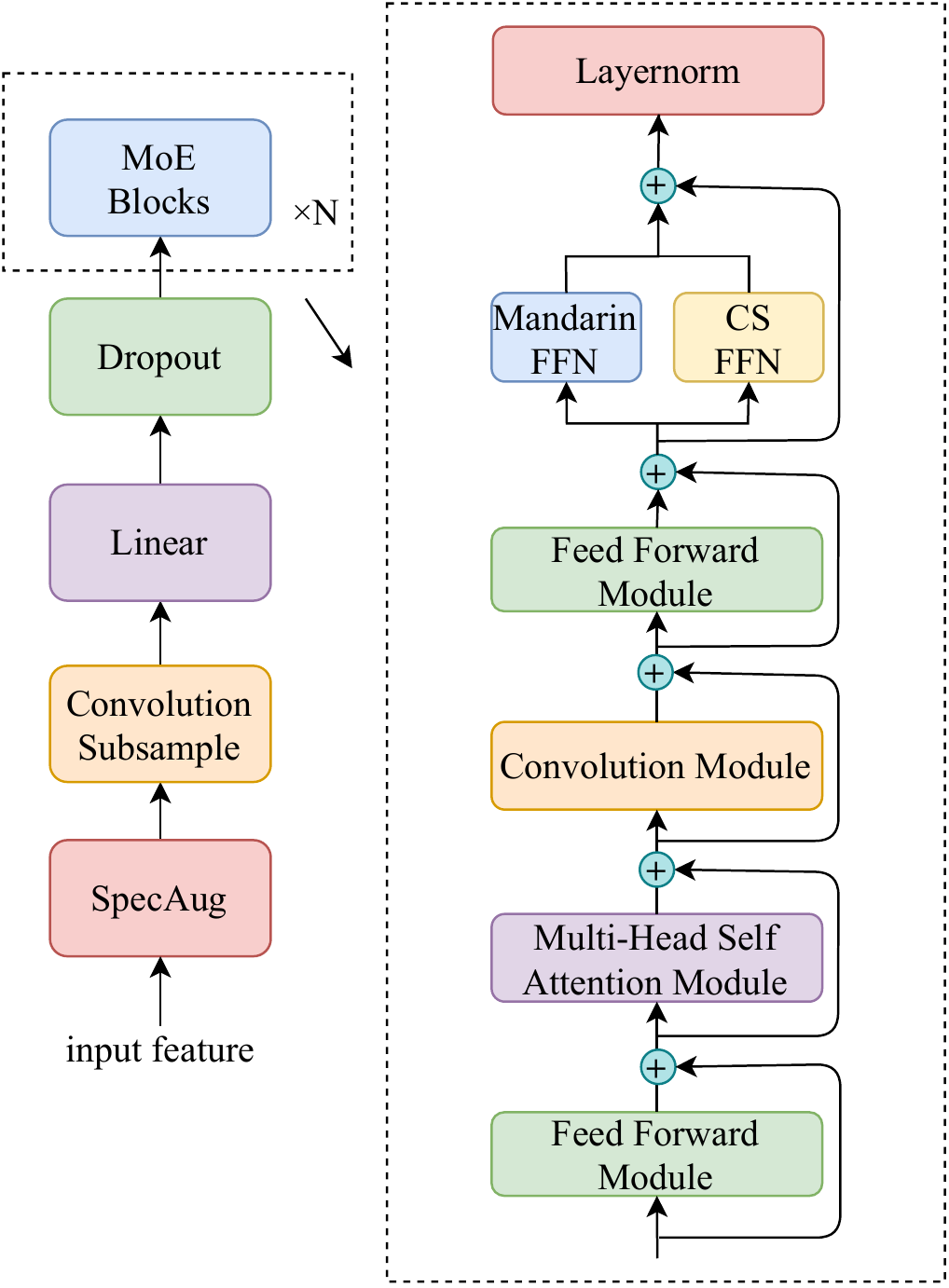}
	\caption{
	The architecture of MoE
	}
	\label{moe}
\vspace{-15pt}
\end{figure}

\vspace{-0.1cm}
Inspired by SpeechMoE~\cite{you2021speechmoe}, we further broaden the shared part in the encoder so that only the last feed-forward layer in each conformer blocks keeps separated. In our implementation of Mixture of Experts, the gating network is not utilized and the output of different experts are added directly. The detail of our MoE structure are shown in Figure \ref{moe}.

\vspace{-0.3cm}
\subsection{Language Model\label{sec32}}
\vspace{-0.1cm}
We train a transformer language model with ESPnet toolkit~\cite{watanabe2018espnet} as our base model. In order to strengthen the contextual modeling capabilities, the following two methods are further investigated.
\vspace{-0.3cm}
\subsubsection{Internal Language Model\label{sec321}}
\vspace{-0.2cm}
At first, shallow fusion~\cite{inproceedings} is used to combine the acoustic model score and language model score. The ASR inference can be described as follow.
\vspace{-0.1cm}
\begin{equation}\label{eq:shallowfusion}
\hat{Y}=\underset{Y}{\operatorname{argmax}}\left(\log P(Y \mid X)+\lambda_{\text{LM}} \log P_{\text{LM}}(Y)\right),
\end{equation}
where $X$ is input acoustic feature, and $Y$ is inferred token sequence, $\log P(Y \mid X)$ is a joint posterior probability, while $\log P_{\text{LM}}(Y)$ is the probability estimated from external language model. 
Like \cite{peng2022internal}, when taking internal LM (ILM)~\cite{peng2022internal} score into consideration, Eq. (\ref{eq:shallowfusion}) is re-formulated as:
\begin{equation}\label{eq:ILM}
\vspace{-0.2cm}
\begin{aligned}
\hat{Y} =&\underset{Y}{\operatorname{argmax}}\left(\log P(Y \mid X)-\lambda_{\text{ILM}} \log P_{\text{ILM}}(Y)\right.\\
&\left.+\lambda_{\text{LM}} \log P_{\text{LM}}(Y)\right). 
\end{aligned}
\end{equation}

ILM casts off higher-level representation from the encoder and only uses the current decoder state to produce the last state. Therefore, the src-attention block is replaced with a feed forward neural network (FFNN) layer. Once the training of the entire AED model is done, we freeze other parameters and continue to train the FFNN block to minimize the perplexity on the training transcripts.
\vspace{-0.3cm}
\subsubsection{Long Context Language Model (LCLM)\label{sec322}}
\vspace{-0.1cm}
Our target is conversational code-switching ASR and leveraging context information has previously proven to be a significant factor in improving the performance of conversational ASR. We implements a simple way to augment code-switching text data for long context language model training~\cite{wei2021context}. Specifically, sentences in the same dialogue are randomly concatenated by pairs in chronological order. Thus our text data for language model training is tripled in this way. The paired sentences are used to train a standard transformer language model which covers longer content from the transcripts.
\vspace{-0.2cm}
\subsection{Training Strategy\label{sec33}}
\vspace{-0.1cm}
To achieve more effective TTS utilization, the ASR model is optimized with an additional consistency loss~\cite{chen2022tts4pretrain} which enforces the posterior distribution similarity between paired real and synthetic speech. The consistency loss is formulated as:
\begin{equation}\label{eq:con-loss}
\vspace{-0.1cm}
\mathcal{L}_{\text {cons }}=\mathcal{D}_{\mathrm{KL}}\left(p_{\theta}(y \mid x) \| p_{\theta}\left(y \mid x^{*}\right)\right) .
\end{equation}
where $\mathcal{D}_{\mathrm{KL}}$ indicates Kullback–Leibler divergence (KL), $x$ is real speech and $y$ is the corresponding transcription.  Synthetic speech $x^{*}$ is generated via TTS of text $y$.
Paired real and synthetic data is required when training with the above consistency loss. With the consistency loss, the standard conformer-based joint CTC/attention loss becomes:
\vspace{-0.2cm}
\begin{equation}
\begin{aligned}
\mathcal{L}=& (1 - \lambda_{1}- \lambda_{2})\left(\mathcal{L}_{\text{att}}\left(Y \mid X \right)+\mathcal{L}_{\text{att}}\left(Y \mid X^{*} \right)\right)\\
& + \lambda_{1}\left( \mathcal{L}_\text{CTC}\left(Y \mid X \right) + \mathcal{L}_\text{CTC}\left(Y \mid X^{*} \right) \right)\\
& + \lambda_{2} \mathcal{L}_{\text {cons}}\left(Y \mid X, X^{*}\right)
\vspace{-0.3cm}
\end{aligned}
\end{equation}
where $\mathcal{L}_{\text {att}}$ is the attention-based cross entropy and $\mathcal{L}_{CTC}$ is the CTC loss.
\vspace{-0.3cm}
\subsection{Model fusion\label{sec34}}
\vspace{-0.1cm}
We fuse the multiple outputs from the above ASR models by ROVER~\cite{fiscus1997post}. The original ROVER is implemented by aligning multiple ASR results and votes with the same weight. We notice that there are many substitution errors on English words replaced by Mandarin characters which have similar pronunciation in code-switching ASR. Then we take unbalanced weights on English words and Mandarin characters during fusion. The weights are empirically selected on the dev set.
\vspace{-0.1cm}
\section{Experiments\label{sec4}}
\subsection{Experiments Setups\label{sec41}}
\vspace{-0.1cm}
All our experiments are conducted on ESPnet~\cite{watanabe2018espnet} and WeNet~\cite{yao2021wenet,zhang2022wenet} toolkits and mix error rate (MER) is measured on the official dev and test sets of the challenge. For the model trained with ESPnet, the acoustic feature of 80 dimensions log Mel-filter bank (Fbank) is extracted from every frame with a frame length of 16ms and frame shift of 8ms. As for WeNet, the window size is 25 ms with a shift of 10ms. 

As for the acoustic model, the standard conformer based on joint CTC is composed of a 12-layer encoder and a 6-layer decoder. The self-attention and the feed-forward sub-layers have 512 and 1024 hidden units, respectively. The number of heads for multi-head attention is 8 in all attention sub-layers. The models trained with WeNet are configured with the same settings as above. For Mandarin, 4340 characters are used as the modeling unit. For English, we use byte pair encoding (BPE) to generate 1000 subwords as the modeling unit. We set the CTC weight to 0.3 both in the training and decoding phase. The language model weight is set to 0.3 during decoding. The beam size is set to 20. We train 40 epochs for the conformer model with a learning rate of 0.0005.

As mentioned in Section \ref{sec31}, besides standard conformer, three variants of encoder are also implemented. Specifically, the bi-encoder consists of 12 Mandarin and English blocks and the LAE consists of 10 shared blocks followed by 10 expert blocks. As for the MoE, we set the number of experts of MoE layers to 3 and it has the same setting as the standard conformer, except for the feed-froward layer dimension modified from 1024 to 2048.
\vspace{-0.2cm}
\begin{table}[h]
\caption{The MER of different augmentation methods}
\setlength{\tabcolsep}{1.0mm}{
\begin{tabular}{cclcl}
\toprule
    Tool            & ID & Model             & dev set(\%)&  \\ \midrule
\multirow{5}{*}{WeNet}  & A1 & Conformer  & 28.38    &  \\
                        & A2 & \quad+ speed perturb.               & \textbf{27.42}    &  \\
                        & A3 & \quad+ pitch shifting               & 27.61    &  \\
                        & A4 & \quad+ speed perturb. \& pitch shifting             & 27.89    &  \\
                        & A5 & \quad+ audio codec               & 28.43    &  \\ \midrule
\multirow{3}{*}{ESPnet} & A6 & Conformer & 23.53    &  \\
                        & A7 & \quad+ synthetic data       & 21.94    &  \\
                        & A8 & \quad\quad+ consistency loss   & \textbf{20.71}    &   \\\bottomrule
\end{tabular}}
\label{tab:result_augment}
\end{table}

\vspace{-0.6cm}
\subsection{Results and analysis\label{sec42}}
\vspace{-0.1cm}
\subsubsection{Comparison of different augmentation methods\label{sec421}}
\vspace{-0.1cm}
To verify the data augmentation methods, experiments are is conducted to the standard conformer. In Table \ref{tab:result_augment}, the typical augmentation methods are examined on the conformers trained using WeNet, while TTS augmentation is studied on the conformers trained using ESPnet. We can see speed perturbation gets better performance compared with other signal pertubation approaches. Pitch shifting reduces the accuracy when working with speed perturbation. Audio codec augmentation unfortuatly leads to negtive efforts. brings few benefits so we did not merge it with other approaches. The use of synthetic data reduce 1.59\% on the MER. Taking benefit from the consistency loss, it brings 1.23\% absolute MER reduction further.

\vspace{-0.2cm}
\subsubsection{Comparison of different acoustic models\label{sec422}}
\vspace{-0.1cm}
Encoder architectures are compared in Table \ref{tab:result_acoustic}. The baseline indicates the result of the official transformer-based baseline. From Table \ref{tab:result_acoustic}, we can see an obvious trend is that more shared parameters lead to better performance. This phenomenon is consistent in both the dev and test sets. We believe that this trend is mainly because the acoustic difference between the two datasets is not big enough and some exclusive parameters are redundant in Bi-encoder and LAE.
\vspace{-0.1cm}
\begin{table}[h]
\caption{The MER on different encoders. Here transformer LM is used with shallow fusion (weight: 0.3) during decoding.}
\setlength{\tabcolsep}{1.8mm}{
\begin{tabular}{lclcc}
\toprule
\multicolumn{1}{l}{}    & ID & Model      & dev set(\%) & test set(\%) \\ \midrule
Baseline                & B1 & Transformer  & 29.20    & 26.50     \\ \midrule
\multirow{3}{*}{WeNet}  & A1 & Conformer  & 28.38    & 27.31     \\
                        & B2 & Bi-encoder & 26.57    & 25.01     \\
                        & B3 & LAE        & \textbf{23.85}    & \textbf{21.80}     \\ \midrule
\multirow{3}{*}{ESPnet} & A6 & Conformer  & 23.53    & 22.23     \\
                        & B4 & LAE        & 21.98    & 21.03     \\
                        & B5 & MoE        & \textbf{21.12}    & \textbf{20.35}     \\ \bottomrule
\end{tabular}}
\label{tab:result_acoustic}
\end{table}

\vspace{-0.6cm}
\subsubsection{Comparison of different language models \label{sec423}}
\vspace{-0.1cm}
We verify ILM and LCLM in Table \ref{tab:result_lm}. The standard transformer language model works better compared to ILM and LCLM in most cases. From the results of B3.1 and B4.1, ILM and LCLM do not show their advantages. But in our further analysis in Section \ref{sec424}, they can produce different views of the hypothesis from the acoustic model and thus the final result will benefit from model fusion. The result of B3.2 shows that ILM is harmful when it is used together with the transformer language model.
\vspace{-0.2cm}
\begin{table}[h]
\caption{The MER on the use of language models. Here LM indicates the transformer-based LM.}
\setlength{\tabcolsep}{0.8mm}{
\begin{tabular}{llllcc}
\toprule
\multicolumn{2}{l}{}        Tool      & ID                  & Model                                     & dev set(\%)    & test set(\%) \\ \midrule
\multicolumn{2}{c}{\multirow{3}{*}{WeNet}}  & B3 & LAE + LM                   & \textbf{23.85}          & \textbf{21.80}        \\
\multicolumn{2}{c}{}                        & B3.1 & LAE + ILM                     & 23.98          & 21.81        \\
\multicolumn{2}{c}{}                        & B3.2 & LAE + LM + ILM & 23.92          & 22.02        \\ \midrule
\multicolumn{2}{c}{\multirow{2}{*}{ESPnet}} & B4 & LAE + LM                  & 21.98 & \textbf{21.00}           \\
\multicolumn{2}{c}{}                        &  B4.1 & LAE + LCLM      & \textbf{21.95}          & 21.04        \\ \bottomrule
\end{tabular}}
\label{tab:result_lm}
\end{table}
\vspace{-0.6cm}
\subsubsection{Model fusion results\label{sec424}}
\vspace{-0.1cm}
We first examine the best single systems before system fusion and the results are reported in Table~\ref{tab:result_last}. Here ESPnet is used for model training. With the help from speed perturbation and training using synthetic data with consistency loss, the LAE and MoE systems can achieve even lower MER on  dev and test sets.

\begin{table}[h]
\caption{The MER of single best systems built on ESPnet. Here sp stands for speed perturbation.}
\setlength{\tabcolsep}{1.5mm}{
\begin{tabular}{clccc}
\toprule
ID & AM                & LM             & dev set(\%) & test set(\%)\\ \midrule
C1 & \begin{tabular}[c]{@{}l@{}}LAE + sp \\ + synthetic data \\ + cons. loss\end{tabular}             & LM & 20.87    & 20.04 \\
C1.1 & \begin{tabular}[c]{@{}l@{}}LAE + sp \\ + synthetic data \\ + cons. loss\end{tabular}             & LCLM & \textbf{20.85}    & \textbf{20.03} \\ \midrule
C2 & \begin{tabular}[c]{@{}l@{}}MoE + sp \\ + synthetic data \\ + cons. loss\end{tabular} & LM & 19.94    & \textbf{19.16}     \\
C2.1 & \begin{tabular}[c]{@{}l@{}}MoE + sp \\ + synthetic data \\ + cons. loss\end{tabular} & LCLM & \textbf{19.91}    & 19.18     \\ \bottomrule
\end{tabular}}
\label{tab:result_last}
\end{table}

\begin{table}[h]
\caption{ROVER based system fusion results with different voting weights on English (EN) words.}
\setlength{\tabcolsep}{2.0mm}{
\begin{tabular}{ccll}
\toprule
Combination                          & EN weight & \multicolumn{1}{c}{dev set(\%)} & \multicolumn{1}{c}{test set(\%)} \\ \midrule
\multirow{5}{*}{\begin{tabular}[l]{@{}l@{}}A3, A8, B2, B3,\\ B4, B5, C1, C2\\ (Submission) \end{tabular}} & 1  & 18.82   &  17.46     \\
                                      & 2         & 18.21     &     16.96                      \\  
                                      & 3         & 18.07     &     16.88                            \\ 
                                      & 4         & \textbf{18.03}     &   \textbf{16.87}       \\
                                      & 5         & 18.20     &    16.93    \\ \midrule
\begin{tabular}[l]{@{}l@{}} A3, A8, B2, B3.1,\\ B4.1, B5, C1, C2.1\\ (Post-challenge) \end{tabular}    & 4 & \textbf{17.81}     & \textbf{16.70}    \\ \bottomrule
\end{tabular}}
\label{tab:result_enweight}
\vspace{-0.1cm}
\end{table}

We finally fuse 8 systems with the ROVER tool and tune the voting weight. Specifically, we set the Mandarin character weight to 1 and adjust the English word weight from 1 to 5 accordingly. Results in Table \ref{tab:result_enweight} show that when the weight of the English word is set four times of that of the Mandarin character, the best performance is achieved. Here 16.87\% is the MER our submission system obtained and the gain from system fusion is significant. After the challenge, we take different language models into consideration and try another combination and this post-challenge system gets a further MER reduction of 0.17\% on the test set.

\section{Conclusion\label{sec5}}
This paper has described our code-switching ASR system for the ISCSLP 2022 Magichub Code-Switching ASR Challenge. In our approach, expert systems and their variants were particularly explored for acoustic modeling.  Among them, MoE shows the best performance with the least language-specific parameters.
As for language modeling, internal LM and long-context LM were studied besides the standard transformer LM. 
Eventually the two language models can reduce MER by producing more candidates for the ROVER tool. The effectiveness of various data augmentation approaches, particularly TTS based augmentation, were investigated as well. We find that additional consistency loss in model training can make better use of the synthetic data and obtain further MER reduction. We obtain extra gain by using ROVER to fuse multiple hypothesis from various models. Compared with the official baseline system, our submission has obtained 11.17\% and 9.63\% absolute MER reduction on the dev and test sets, respectively. 

\newpage

\bibliographystyle{IEEEtran}

\bibliography{template}

\begin{thebibliography}{10}
\providecommand{\url}[1]{#1}
\csname url@samestyle\endcsname
\providecommand{\newblock}{\relax}
\providecommand{\bibinfo}[2]{#2}
\providecommand{\BIBentrySTDinterwordspacing}{\spaceskip=0pt\relax}
\providecommand{\BIBentryALTinterwordstretchfactor}{4}
\providecommand{\BIBentryALTinterwordspacing}{\spaceskip=\fontdimen2\font plus
\BIBentryALTinterwordstretchfactor\fontdimen3\font minus
  \fontdimen4\font\relax}
\providecommand{\BIBforeignlanguage}[2]{{%
\expandafter\ifx\csname l@#1\endcsname\relax
\typeout{** WARNING: IEEEtran.bst: No hyphenation pattern has been}%
\typeout{** loaded for the language `#1'. Using the pattern for}%
\typeout{** the default language instead.}%
\else
\language=\csname l@#1\endcsname
\fi
#2}}
\providecommand{\BIBdecl}{\relax}
\BIBdecl

\bibitem{guo2018study}
P.~Guo, H.~Xu, L.~Xie, and E.~S. Chng, ``Study of semi-supervised approaches to
  improving english-mandarin code-switching speech recognition,'' in
  \emph{Proc. INTERSPEECH}.\hskip 1em plus 0.5em minus 0.4em\relax ISCA, 2018,
  pp. 1928--1932.

\bibitem{shi2020asru}
X.~Shi, Q.~Feng, and L.~Xie, ``The {ASRU} 2019 mandarin-english code-switching
  speech recognition challenge: Open datasets, tracks, methods and results,''
  in \emph{Proc. ASRU}, 2020.

\bibitem{shah2020first}
S.~Shah, S.~Sitaram, and R.~Mehta, ``First workshop on speech processing for
  code-switching in multilingual communities: Shared task on code-switched
  spoken language identification,'' in \emph{Proc. WSTCSMC}, 2020, p.~24.

\bibitem{diwan2021multilingual}
A.~Diwan, R.~Vaideeswaran, S.~Shah, A.~Singh, S.~R.~K. M., S.~Khare, V.~Unni,
  S.~Vyas, A.~Rajpuria, C.~Yarra, A.~R. Mittal, P.~K. Ghosh, P.~Jyothi,
  K.~Bali, V.~Seshadri, S.~Sitaram, S.~Bharadwaj, J.~Nanavati, R.~Nanavati, and
  K.~Sankaranarayanan, ``{MUCS} 2021: Multilingual and code-switching {ASR}
  challenges for low resource indian languages,'' in \emph{Proc.
  INTERSPEECH}.\hskip 1em plus 0.5em minus 0.4em\relax ISCA, 2021, pp.
  2446--2450.

\bibitem{vaswani2017attention}
A.~Vaswani, N.~Shazeer, N.~Parmar, J.~Uszkoreit, L.~Jones, A.~N. Gomez,
  {\L}.~Kaiser, and I.~Polosukhin, ``Attention is all you need,'' in
  \emph{Proc. NIPS}, 2017.

\bibitem{chorowski2015attention}
J.~Chorowski, D.~Bahdanau, D.~Serdyuk, K.~Cho, and Y.~Bengio, ``Attention-based
  models for speech recognition,'' in \emph{Proc. NIPS}, 2015, pp. 577--585.

\bibitem{graves2012sequence}
A.~Graves, ``Sequence transduction with recurrent neural networks,'' in
  \emph{Proc. ICML}, 2012.

\bibitem{lu2020bi}
Y.~Lu, M.~Huang, H.~Li, J.~Guo, and Y.~Qian, ``Bi-encoder transformer network
  for mandarin-english code-switching speech recognition using mixture of
  experts,'' in \emph{Proc. INTERSPEECH}.\hskip 1em plus 0.5em minus
  0.4em\relax ISCA, 2020, pp. 4766--4770.

\bibitem{tian2022lae}
J.~Tian, J.~Yu, C.~Zhang, C.~Weng, Y.~Zou, and D.~Yu, ``Lae: Language-aware
  encoder for monolingual and multilingual asr,'' in \emph{Proc.
  INTERSPEECH}.\hskip 1em plus 0.5em minus 0.4em\relax ISCA, 2022.

\bibitem{hou2021exploiting}
W.~Hou, H.~Zhu, Y.~Wang, J.~Wang, T.~Qin, R.~Xu, and T.~Shinozaki, ``Exploiting
  adapters for cross-lingual low-resource speech recognition,'' in \emph{Proc.
  TASLP}, 2022, pp. 317--329.

\bibitem{chen2022tts4pretrain}
Z.~Chen, Y.~Zhang, A.~Rosenberg, B.~Ramabhadran, P.~J. Moreno, and G.~Wang,
  ``Tts4pretrain 2.0: Advancing the use of text and speech in {ASR} pretraining
  with consistency and contrastive losses,'' in \emph{Proc. ICASSP}.\hskip 1em
  plus 0.5em minus 0.4em\relax IEEE, 2022, pp. 7677--7681.

\bibitem{chen2021injecting}
Z.~Chen, Y.~Zhang, A.~Rosenberg, B.~Ramabhadran, G.~Wang, and P.~J. Moreno,
  ``Injecting text in self-supervised speech pretraining,'' in \emph{Proc.
  ASRU}.\hskip 1em plus 0.5em minus 0.4em\relax IEEE, 2021, pp. 251--258.

\bibitem{park2022unsupervised}
C.~Park, R.~Ahmad, and T.~Hain, ``Unsupervised data selection for speech
  recognition with contrastive loss ratios,'' in \emph{Proc. ICASSP}.\hskip 1em
  plus 0.5em minus 0.4em\relax IEEE, 2022, pp. 8587--8591.

\bibitem{hu2022synt++}
T.~Hu, M.~Armandpour, A.~Shrivastava, J.~R. Chang, H.~Koppula, and O.~Tuzel,
  ``{SYNT++:} utilizing imperfect synthetic data to improve speech
  recognition,'' in \emph{Proc. ICASSP}.\hskip 1em plus 0.5em minus 0.4em\relax
  IEEE, 2022, pp. 7682--7686.

\bibitem{liu2021code}
G.~Liu and L.~Cao, ``Code-switch speech rescoring with monolingual data,'' in
  \emph{Proc. ICASSP}.\hskip 1em plus 0.5em minus 0.4em\relax IEEE, 2021, pp.
  6229--6233.

\bibitem{watanabe2018espnet}
S.~Watanabe, T.~Hori, S.~Karita, T.~Hayashi, J.~Nishitoba, Y.~Unno, N.~E.~Y.
  Soplin, J.~Heymann, M.~Wiesner, N.~Chen, A.~Renduchintala, and T.~Ochiai,
  ``Espnet: End-to-end speech processing toolkit,'' in \emph{Proc.
  INTERSPEECH}.\hskip 1em plus 0.5em minus 0.4em\relax ISCA, 2018, pp.
  2207--2211.

\bibitem{yao2021wenet}
Z.~Yao, D.~Wu, X.~Wang, B.~Zhang, F.~Yu, C.~Yang, Z.~Peng, X.~Chen, L.~Xie, and
  X.~Lei, ``Wenet: Production oriented streaming and non-streaming end-to-end
  speech recognition toolkit,'' in \emph{Proc. INTERSPEECH}.\hskip 1em plus
  0.5em minus 0.4em\relax ISCA, 2021, pp. 4054--4058.

\bibitem{wei2021context}
K.~Wei, P.~Guo, H.~Lv, Z.~Tu, and L.~Xie, ``Context-aware rnnlm rescoring for
  conversational speech recognition,'' in \emph{Proc. ISCSLP}.\hskip 1em plus
  0.5em minus 0.4em\relax IEEE, 2021, pp. 1--5.

\bibitem{fiscus1997post}
J.~G. Fiscus, ``A post-processing system to yield reduced word error rates:
  Recognizer output voting error reduction (rover),'' in \emph{Proc.
  ASRU}.\hskip 1em plus 0.5em minus 0.4em\relax IEEE, 1997, pp. 347--354.

\bibitem{yu2022summary}
F.~Yu, S.~Zhang, P.~Guo, Y.~Fu, Z.~Du, S.~Zheng, W.~Huang, L.~Xie, Z.~Tan,
  D.~Wang, Y.~Qian, K.~A. Lee, Z.~Yan, B.~Ma, X.~Xu, and H.~Bu, ``Summary on
  the {ICASSP} 2022 multi-channel multi-party meeting transcription grand
  challenge,'' in \emph{Proc. ICASSP}.\hskip 1em plus 0.5em minus 0.4em\relax
  IEEE, 2022, pp. 9156--9160.

\bibitem{sun2018multiple}
S.~Sun, Y.~Shi, C.-F. Yeh, S.~Bu, M.-Y. Hwang, and L.~Xie, ``Multiple
  beamformers with rover for the chime-5 challenge,'' in \emph{Proc. 5th
  International Workshop on Speech Processing in Everyday Environments}, 2018.

\bibitem{wang2022sjtu}
W.~Wang, X.~Gong, Y.~Wu, Z.~Zhou, C.~Li, W.~Zhang, B.~Han, and Y.~Qian, ``The
  sjtu system for multimodal information based speech processing challenge
  2021,'' in \emph{Proc. ICASSP}.\hskip 1em plus 0.5em minus 0.4em\relax IEEE,
  2022, pp. 9261--9265.

\bibitem{li2022talcs}
C.~Li, S.~Deng, Y.~Wang, G.~Wang, Y.~Gong, C.~Chen, and J.~Bai, ``{TALCS:} an
  open-source mandarin-english code-switching corpus and a speech recognition
  baseline,'' in \emph{Proc. INTERSPEECH}.\hskip 1em plus 0.5em minus
  0.4em\relax ISCA, 2022.

\bibitem{yang2022open}
Z.~Yang, Y.~Chen, L.~Luo, R.~Yang, L.~Ye, G.~Cheng, J.~Xu, Y.~Jin, Q.~Zhang,
  P.~Zhang \emph{et~al.}, ``Open source magicdata-ramc: A rich annotated
  mandarin conversational (ramc) speech dataset,'' in \emph{Proc.
  INTERSPEECH}.\hskip 1em plus 0.5em minus 0.4em\relax ISCA, 2022.

\bibitem{park2019specaugment}
K.~Wei, P.~Guo, H.~Lv, Z.~Tu, and L.~Xie, ``Context-aware {RNNLM} rescoring for
  conversational speech recognition,'' in \emph{Proc. ISCSLP}.\hskip 1em plus
  0.5em minus 0.4em\relax IEEE, 2021, pp. 1--5.

\bibitem{Liu2021DelightfulTTS}
Y.~Liu, Z.~Xu, G.~Wang, K.~Chen, B.~Li, X.~Tan, J.~Li, L.~He, and S.~Zhao,
  ``Delightfultts: The microsoft speech synthesis system for blizzard challenge
  2021,'' in \emph{arXiv preprint arXiv:2110.12612}, 2021.

\bibitem{Kong2020HiFiGANGA}
J.~Kong, J.~Kim, and J.~Bae, ``Hifi-gan: Generative adversarial networks for
  efficient and high fidelity speech synthesis,'' in \emph{Proc. NeurIPS 2020},
  2020.

\bibitem{Wang2018StyleTU}
Y.~Wang, D.~Stanton, Y.~Zhang, R.~J. Skerry{-}Ryan, E.~Battenberg, J.~Shor,
  Y.~Xiao, Y.~Jia, F.~Ren, and R.~A. Saurous, ``Style tokens: Unsupervised
  style modeling, control and transfer in end-to-end speech synthesis,'' in
  \emph{Proc. ICML}.\hskip 1em plus 0.5em minus 0.4em\relax PMLR, 2018, pp.
  5167--5176.

\bibitem{you2021speechmoe}
Z.~You, S.~Feng, D.~Su, and D.~Yu, ``Speechmoe: Scaling to large acoustic
  models with dynamic routing mixture of experts,'' in \emph{Proc.
  INTERSPEECH}.\hskip 1em plus 0.5em minus 0.4em\relax ISCA, 2021, pp.
  2077--2081.

\bibitem{inproceedings}
D.~Zhao, T.~N. Sainath, D.~Rybach, P.~Rondon, D.~Bhatia, B.~Li, and R.~Pang,
  ``Shallow-fusion end-to-end contextual biasing,'' in \emph{Proc.
  INTERSPEECH}.\hskip 1em plus 0.5em minus 0.4em\relax ISCA, 2019, pp.
  1418--1422.

\bibitem{peng2022internal}
Y.~Peng, Y.~Liu, J.~Zhang, H.~Xu, Y.~He, H.~Huang, and E.~S. Chng, ``Internal
  language model estimation based language model fusion for cross-domain
  code-switching speech recognition,'' in \emph{arXiv preprint
  arXiv:2207.04176}, 2022.

\bibitem{zhang2022wenet}
B.~Zhang, D.~Wu, Z.~Peng, X.~Song, Z.~Yao, H.~Lv, L.~Xie, C.~Yang, F.~Pan, and
  J.~Niu, ``Wenet 2.0: More productive end-to-end speech recognition toolkit,''
  in \emph{arXiv preprint arXiv:2203.15455}, 2022.

\end{thebibliography}

\end{document}